\documentclass[aps,twocolumn,showpacs,preprintnumbers,prb,superscriptaddress]{revtex4}

\usepackage{graphicx}
\usepackage{amssymb}
\usepackage{amsmath}
\usepackage{color}
\usepackage[toc,page]{appendix}

\def\redu{\eta^u}
\def\sredu{x^u}
\def\redub{\upsilon^u}
\newcommand{\R}{\mathbf{r}}
\newcommand{\UP}{n_{\uparrow}}
\newcommand{\DN}{n_{\downarrow}}

\begin{document}

\title{Hartree potential dependent exchange functional}
\author{Lucian A. Constantin}
\affiliation{Center for Biomolecular Nanotechnologies @UNILE, Istituto Italiano di Tecnologia, Via Barsanti, I-73010 
Arnesano, Italy}
\author{Eduardo Fabiano}
\affiliation{Istituto Nanoscienze-CNR, Euromediterranean Center for Nanomaterial Modelling and Technology (ECMT), via 
Arnesano, Lecce 73100, Italy}
\affiliation{Center for Biomolecular Nanotechnologies @UNILE, Istituto Italiano di Tecnologia, Via Barsanti, I-73010 
Arnesano, Italy}
\author{Fabio Della Sala}
\affiliation{Istituto Nanoscienze-CNR, Euromediterranean Center for Nanomaterial Modelling and Technology (ECMT), via 
Arnesano, Lecce 73100, Italy}
\affiliation{Center for Biomolecular Nanotechnologies @UNILE, Istituto Italiano di Tecnologia, Via Barsanti, I-73010 
Arnesano, Italy}

\date{\today}

\begin{abstract}
We introduce a novel non-local ingredient for the construction of exchange density functionals:
the reduced Hartree parameter, which is invariant under the uniform 
scaling of the density and represents the exact exchange enhancement factor for  one- and two-electron systems.
The reduced Hartree parameter is used together with the conventional
meta-generalized gradient approximation (meta-GGA) semilocal ingredients  (i.e. the electron density, its gradient 
and the kinetic energy density) to construct a new generation exchange functional, termed u-meta-GGA.
This u-meta-GGA functional is exact for {the exchange of} any one- and two-electron systems, 
is size-consistent and non-empirical, satisfies the uniform density scaling relation, and 
recovers the modified gradient expansion derived from the semiclassical atom theory.
For atoms, ions, jellium spheres, and molecules, it shows a good accuracy, 
being often better than meta-GGA exchange functionals.
Our construction validates the use of the reduced Hartree ingredient in exchange-correlation 
functional development, opening the way to an additional rung in the Jacob's ladder classification 
of non-empirical density functionals.
 
\end{abstract}


\maketitle

\section{Introduction}

Kohn-Sham (KS) ground-state density functional theory (DFT) 
\cite{kohnPR1965,dobson_vignale_book,yang_paar_book,seminario_book,sholl_book,jonesRMP2015,burkeJCP2012}
is one the most used methods in electronic calculations of quantum
chemistry and condensed-matter physics. Its practical implementation
is based on approximations of the exchange-correlation (XC) energy ($E_{xc}$),
which is a subject of intense research \cite{scuseriaREVIEW05,jonesRMP2015,burkeJCP2012}.

The simplest functionals, beyond the local density approximation 
\cite{kohnPR1965} (LDA),
are  those based on the generalized gradient approximation (GGA), which are
constructed using the electron density ($n$) and 
its reduced gradients (e.g. $s$ in Eq. (\ref{eq9b})).
These functionals can achieve reasonable accuracy for various energetical 
and/or structural properties of molecules and/or solids, at 
a moderate computational cost \cite{langrethPRB1983,perdewPRL96,perdewPRL08,constantinPRL11,fabianoJCTC2011,
peveratiJCTC2012,zhaoJCP08,tognettiCPL08,tognettiJCP08,beckePRA88,leePRB88,carmonaJCP15,armientoPRB05,
fabianoPRB10,constantinPRB11,swartMP04,wilsonIJQC98,thakkarJCP09,fabianoJCTC14gap,constantinJCP12,
chiodoPRL12,constantinPRB16}.
However, because of their simplicity, GGA functionals also show several 
important limitations, especially in terms 
of broad applicability. Moreover, they are based on a heavy error 
cancellation betweeen exchange 
and correlation parts \cite{constantinPRB12}.  

To improve over GGAs, meta-generalized-gradient-approximations (meta-GGAs) can
be considered
\cite{dellasalaARXIV16,perdewPRL99,delCPL12,taoPRL03,perdewPRL09,perdewPRL11,constantinJCTC13,constantin2016semilocal,
sunPRL13,sunJCP13,sunJCP12,ruzsinszkyJCTC12,sunPRL15,sunPNAS15,wellendorffJCP14,zhao2008m06,
peverati2011m11,peveratiPTRSLA14,beckePRA89,zhaoACR2008,peveratiPCCP2012,beckeJCP98,constantinPRB13}.
These are the most sophisticated semilocal functionals and use, 
as additional ingredient with respect to the GGA ones, the
positive-defined kinetic energy density
$\tau=(1/2)\sum_{i=1}^N|\nabla\phi_i|^2$ (with $\phi_i$ being the KS orbitals and
$N$ being the number of occupied KS orbitals).
This quantity enters in the expansion of the angle-averaged
exact exchange hole \cite{beckeIJQC83}, being thus a natural and important 
tool in the construction of XC approximations.
Meta-GGA functionals incorporate important exact conditions and 
have an improved overall accuracy with respect to the GGA functionals.
Moreover, because the kinetic energy density
can be easily computed at any step of the 
KS self-consistent scheme, 
the meta-GGA functionals have almost the same attractive computational 
cost as any GGA. 

Further improvements, beyond the meta-GGA level of theory, are 
usually realized abandoning the semilocal framework. 
Here we mention the so-called 3.5 Rung functionals 
\cite{janesko2012nonspherical,janesko2013rung,janesko2010rung,janesko2012nonempirical}, 
that incorporate a linear dependence on the nonlocal 
one-particle density matrix, and non-local functionals based 
on the properties and modelling of the 
exchange-correlation hole 
\cite{gunnarsson1976exchange,gunnarsson1977exchange,alonso1978nonlocal,gunnarsson1979descriptions,wu2004comparing,giesbertzPRA13}.
Moreover, popular
tools in computational chemistry
are the hybrid functionals
\cite{perdewPRL96_pbe0,burkeCPL1997,marsmanJP08,zhaoJPCA04,beckeJCP93,beckeJCP93_hybrid,baerARPC10,
fabianoJCTC15,fabianoIJQC13}, which mix a fraction of non-local
Hartree-Fock exchange with a semilocal XC functional. Alternatively,
even more complex possibilities can be considered, 
such as hyper-GGA functionals \cite{perdewJCP05,perdewPRA08,haunschildJCP12,beckeJCP07} or
orbital-dependent functionals \cite{kummelRMP08,bartlettJCP05,grabowskiPRB13,grabowskiJCP14}.
In this way, a significant increase of the accuracy can be achieved. 
Nevertheless, because of the need to compute non-local contributions
(e.g. the Hartree-Fock exchange), the computational cost of such methods is
considerably larger than the one of semilocal functionals. 

In this paper, we consider an alternative strategy to introduce non-local
effects into a density functional, without affecting too much the
final computational cost. The idea is to consider, as additional ingredient
beyond the conventional meta-GGA level of theory, the 
Hartree potential
\begin{equation}\label{e1}
u(\R)=\int d\R'\frac{n(\R')}{|\R-\R'|}\ .
\end{equation}
The Hartee potential appears to be a natural input ingredient in 
the construction of exchange functionals for several reasons:
\begin{itemize}
\item For one- and two-electron systems, the exact exchange energy is \cite{yang_paar_book,seminario_book}
\begin{eqnarray}
\label{e2}
E_x[n]=-\frac{1}{2}\int d\R\; n(\R)\;u(\R), \;\;\;\mathrm{for}\;\;\; N=1\ , \\
\label{e3}
E_x[n]=-\frac{1}{4}\int d\R\; n(\R)\;u(\R), \;\;\;\mathrm{for}\;\;\; N=2\ ,
\end{eqnarray}
where $N$ is the number of electrons.
Note that Eq. (\ref{e2}) is the basis of the 
self-interaction correction approach of Perdew and Zunger \cite{perdew1981self}.
\item The asymptotic decay of the
Hartree potential 
\begin{equation}
\lim_{r\rightarrow\infty }u(\R)=N/r 
\label{e4}
\end{equation}  
is proportional to  that of the exact exchange per particle and potential \cite{ayers2005fermi}:
\begin{eqnarray}
 \lim_{r\rightarrow\infty }\epsilon_x(\R)&=&-1/(2r) \\ 
 \lim_{r\rightarrow\infty }v_x(\R)       &=&-1/r.
\end{eqnarray}
In fact the Fermi-Amaldi potential \cite{fa34,parr95} which equals $u(\R)/N$, has been largely used to 
construct exchange and exchange-correlation functionals, see e.g. Refs. \onlinecite{cedillo86,yangwu02,ayers2005,umezawa06}.
However, the Fermi-Amaldi potential depends on $N$, thus it is not size-consistent \cite{TRICKEY2013}.

\end{itemize}
In this work we consider the 
construction of an exchange functional of the general form
\begin{eqnarray}\label{e5}
E_x^{u-MGGA}[n] & = & \int n\epsilon_x^{u-MGGA}(n,\nabla n,\tau,u)d\R = \\
\nonumber
& = & \int n\epsilon_x^{LDA}(n)F_x(n,\nabla n,\tau,u)d\R\ ,
\end{eqnarray}
where $\epsilon_x^{LDA}=-(3/4\pi)(3\pi^2)^{1/3}n^{1/3}$ is the local density approximation for exchange
and $F_x$ is the exchange enhancement factor.
The functional of Eq. (\ref{e5}) constitutes the prototype for a new class of functionals,
that we name {\it u-meta-GGA} (u-MGGA in short).
The construction of a correlation u-meta-GGA functional is also conceivable
but it is a more complex task and it is left for future work.
The u-meta-GGA exchange functional is expected to have higher accuracy than
conventional meta-GGAs, thanks to the inclusion of non-local effects via the
Hartree potential. 
At the same time, because the Hartree potential must be
anyway computed at every step of any KS calculation (even at the LDA level),
it bears no essential additional computational cost with respect to 
meta-GGAs.

\section{Construction of the u-meta-GGA exchange functional}

\subsection{The reduced Hartree ingredient}
\label{x_sect}
To start our work, we consider the construction of a proper
reduced ingredient that depends on the Hartree potential and
has the correct features to be usefully employed in the
construction of density functionals. This is 
the {\it Hartree reduced parameter}
\begin{equation}
\redu=\frac{u}{3(3/\pi)^{1/3}n^{1/3}}\ .
\label{e6}
\end{equation}
This ingredient is invariant under the uniform 
scaling of the density ($n_\gamma(\R)=\gamma^3n(\gamma\R)$,
with $\gamma>0$), i.e. it behaves as $\redu_\gamma(\R)=\redu(\gamma\R)$.
This invariance is a key property for any input ingredient to be used in
the development of semilocal DFT functionals. In fact, it is satisfied by 
all the semilocal ingredients:
\begin{equation}
s=\frac{|\nabla n|}{2 k_F n}\;\; ,\;\; z=\frac{\tau^W}{\tau}\;\;,\;\; \alpha=\frac{\tau-\tau^W}{\tau^{unif}}\ ,
\label{eq9b}
\end{equation}
where $k_F=(3\pi^2n)^{1/3}$ is the Fermi wavevector,
$\tau^{unif}=\frac{3}{10}(3\pi^{2})^{2/3}n^{5/3}$ is the Thomas-Fermi kinetic energy density 
\cite{thomasMPCPS27,fermiRANL27},
and $\tau^W=\tau^{unif}5s^2/3$ is the
von Weizs\"{a}cker kinetic energy density \cite{weizsacker1935theorie}.

Additional important formal properties of $\redu$ can be obtained considering its behavior under
the coordinate and particle-number density scaling \cite{scaling}
($n^{(\beta)}_\gamma(\R)=\gamma^{3\beta+1}n(\gamma^\beta\R)$,
with $\gamma>0$ and $\beta$ being a parameter), which defines
a whole family of scaling relations.
Under this scaling, we have $\redu\rightarrow\gamma^{2/3}\redu$,
i.e. the Hartree reduced ingredient scales as
$N^{2/3}$, with $N$ being the number of electrons.
This result indicates that, unlike the semilocal parameters,
$\redu$ is a size-extensive quantity, 
increasing with the number of electrons. 
We note that, in spite of this feature, 
the reduced Hartree parameter is anyway behaving in a proper size-consistent
way as shown in Appendix  \ref{appsize}.


Moreover, we can note that the scaling properties of
$\redu$ do not depend on the value of the parameter $\beta$.
Thus, for the uniform-electron-gas and the Thomas-Fermi scalings
\cite{scaling,elliottPRL08}, where the $\gamma\rightarrow\infty$ limit
is important, large values of $\redu$ are relevant; on the opposite,
for the homogeneous and fractional-particle scalings \cite{scaling,borgooJCP12,borgooJCTC13},
where $\gamma\rightarrow0$, small values of $\redu$ are important.
These considerations will be significant to analyze
the behavior of an $\redu$-dependent enhancement factor in
different conditions.

\begin{figure*}
\includegraphics[width=0.95\textwidth]{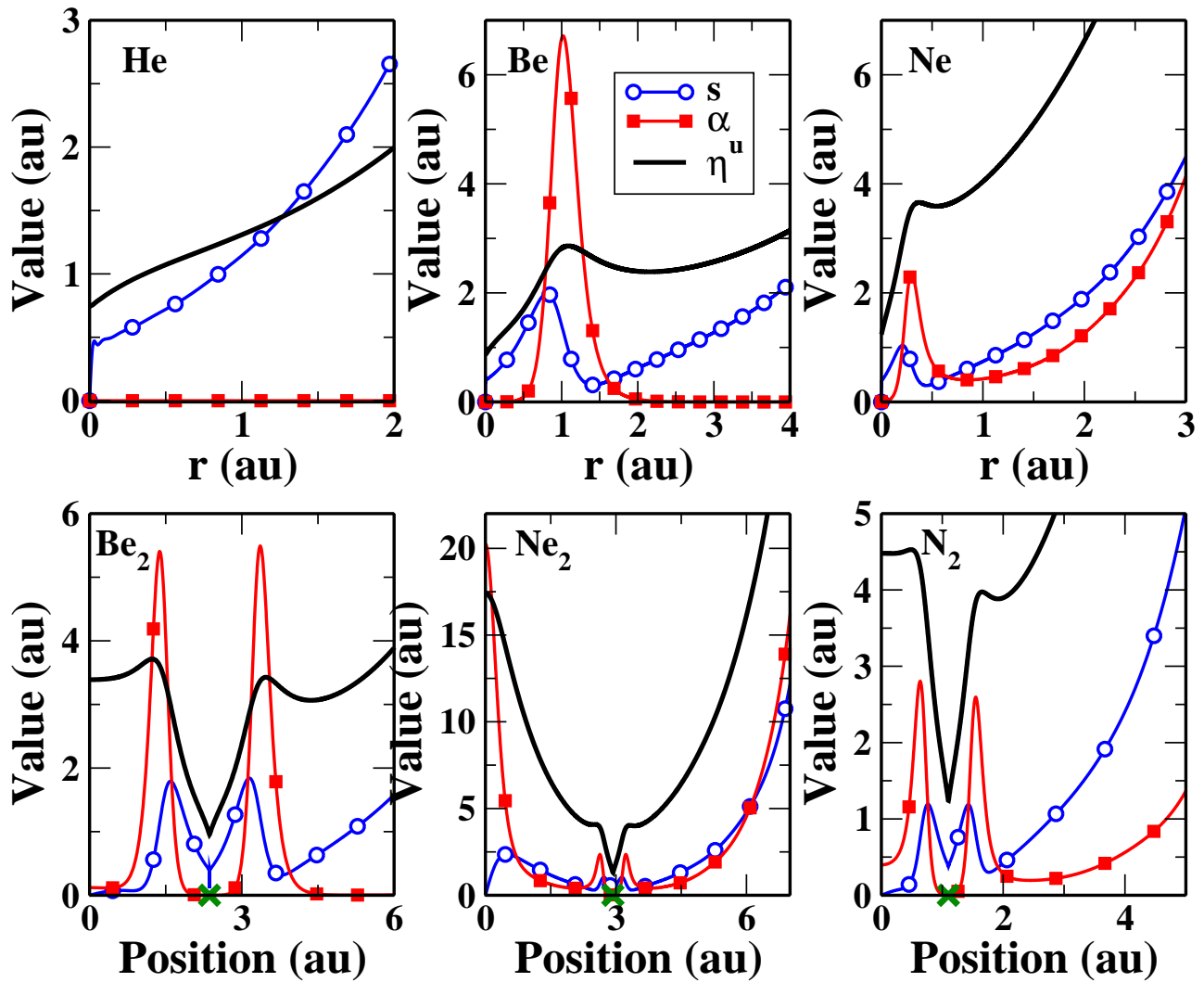}
\caption{\label{fig1}Plot of the reduced gradient ($s$), the $\alpha$ meta-GGA ingredient, and 
the Hartree reduced parameter $\redu$, 
for some atoms (top) and dimers (bottom), as functions of the position. 
The green cross in the bottom part of each dimer plot denotes the position of the 
atom (note that only half of the dimer is plotted).}
\end{figure*}

In Fig. \ref{fig1}, we compare the behavior of $\redu$
with that of the other semilocal ingredients for 
some atoms and dimers.
It can be seen that $\redu$ behaves rather different
than the other conventional semilocal reduced parameters ($s$, $z$,
and $\alpha$), being in general more shallowed and averaged: $\redu$ 
is in fact a non-local ingredient and thus
it contains, at every point of space, information on the whole system.
Moreover, 
 {\it it is larger than zero at any point in space} (unlike $s$, for example); in the density tail asymptotic region we always have  
\begin{equation}
\redu \rightarrow +\infty \; ,
\end{equation}
like $s$ but in contrast to $\alpha$ which vanishes for iso-orbital density tails (e.g. for Be).

Thus, $\redu$ appears to be an 
interesting tool for the construction of advanced functionals
both to complement the information available from standard semilocal 
reduced ingredient and to add information on the
shape of the exchange enhancement factor. 

The most important feature of $\redu$ is that the simple exchange enhancement factor 
\begin{equation}
\label{exact}
F_x=\redu
\end{equation}
yields immediately Eqs. (\ref{e2}) and (\ref{e3}) [for the former,
note that the exchange energy satisfies the spin-scaling relation \cite{oliverPRA79}
$E_x[\UP,\DN]=(E_x[2\UP]+E_x[2\DN])/2$]. 
Hence, $\redu$ {\it represents the exact exchange enhancement factor for any one- and two-electron system}.

Finally, it is also useful to define the bounded ingredient
\begin{equation}
\redub=\frac{1}{1+\redu},
\label{ebi}
\end{equation}
that is small everywhere for large systems, but also in the
 tail of the density (where $\redu\rightarrow\infty$).
%
\begin{figure}
\includegraphics[width=\columnwidth]{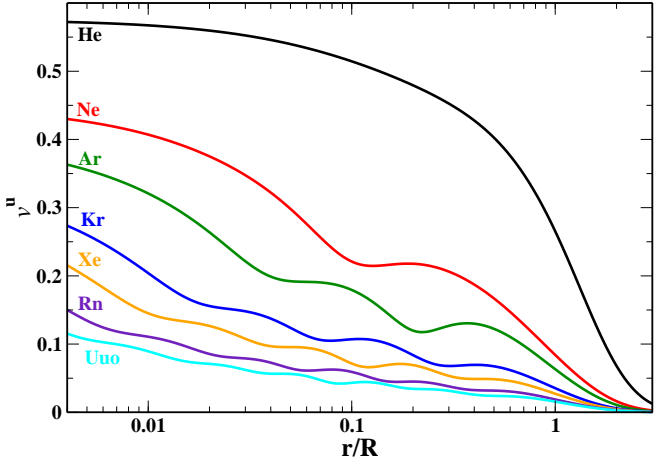}
\caption{The bounded ingredient $\redub$ versus the scaled radial distance $r/R$ for noble atoms (He-Uuo).
Here $R$ is the atomic radius. For He-Rn we use the atomic radii of Ref. \cite{mantina2009consistent}, while for 
Uuo we extrapolate the data of Ref. \cite{mantina2009consistent}, finding $R=2.22 \AA$.
}
\label{f4bbb}
\end{figure}
In Fig. \ref{f4bbb}, we show $\redub$ for the noble atoms of the periodic table. 
One can see that in most of the space the 
curves are not intersecting, such that $\redub$ can be considered a good atomic indicator, 
being of interest for functional development.

\subsection{The u-meta-GGA exchange functional}
\label{umgga_sec}
In the previous subsection we introduced the Hartree reduced parameter
and we showed that it posses interesting properties that
suggest its utility as input quantity in the construction of
advanced density functionals. On the other hand, we have observed that 
in general $\redu$ is always large in magnitude. Thus, the proper use
of this quantity in functional development is not trivial and
the construction of a good u-meta-GGA functional represents instead
a challenge.

To attempt to fulfill this task, 
we consider the following ansatz for 
the u-meta-GGA exchange enhancement factor
\begin{equation}
F_x^{u-MGGA}
= A\; 
F_x^{1}
\label{e7}
\end{equation}
where 

\begin{eqnarray}
A
&=&\frac{\beta +\redu}{1+\beta^{1/\redu}\redu} \\
\label{e8}
\beta & = & \frac{b}{\sqrt{1+s^6}} \\
b&=&(1-z^3)^{a_1}\ ,\\
\label{e9}
F_x^{1} & = & \frac{1+b\left[\mu\frac{3}{5}z
+a_3 s^4 \frac{2\pi}{3\sqrt{5}}\sqrt{\alpha} \right]} {1+ b
a_3  s^4 \sqrt{\ln(1+\alpha)}}\ ,\\
\mu & = & \mu^{MGE2}+a_2 \redub, \label{eqlast}
\end{eqnarray}
with $\mu^{MGE2}=0.26$ being the coefficient of the modified second-order gradient expansion (MGE2) 
\cite{elliottPRL08,elliottCJC09,constantinPRB16}, and 
the  $a_1=1/6$, $a_2=0.05$, and $a_3=0.08$ being non-empirical parameters
fitted to a class of four-electron model systems described in Section \ref{sec:param}.

The function $b=b(z)$ controls the transition from
pure u-meta-GGA behavior ($F_x^{u-MGGA}=\redu$), which is
exact for iso-orbital regions ($z=1$), to a
meta-GGA like behavior $(F_x^{u-MGGA}=F_x^{1}$),
which is appropriate for slowly-varying density 
limit ($z\approx0$).
Moreover, $a_2$ is a parameter which helps to tune
the value of the second-order coefficient in the Taylor expansion at
slowly-varying densities for each atom. This is done using the
parameter $\redub$ as an atomic indicator. Note that for small atoms
$\mu>\mu^{MGE2}$
(here the gradient expansion is less meaningful, and the
results are very sensitive to the functional form)
, whereas in the semiclassical limit (with an infinite number
of electrons) $\mu=\mu^{MGE2}$.
Finally, $a_3$ modulates the behavior of the functional
in the tail of the density.

The u-meta-GGA exchange functional has been costructed satisying the following properties:
\begin{itemize}
\item[-] For one and two electron systems $z=1$, so that $b=\beta=0$, yielding $A=\redu$ and $F_X^1=1$, 
therefore $F_x^{u-MGGA}=\redu$ is exact (see Eq. (\ref{exact}));
\item[-] Under the uniform density scaling $n_\gamma(\R)=\gamma^3n(\gamma\R)$,
with $\gamma>0$, it behaves correctly as $E_x^{u-MGGA}[n_\gamma]=\gamma E_x^{u-MGGA}[n]$; 
\item[-] It is size-consistent (see Appendix \ref{appsize});
\item[-] For many-electron systems, we can distinguish different regions:
\begin{itemize}
\item[--]  In the slowly-varying density limit ($s\rightarrow0$, $z\rightarrow 5s^2/3+\mathcal{O}(|\nabla n|^4)$, $\alpha\rightarrow 
1+\mathcal{O}(|\nabla n|^2)$) 
we have  $b\rightarrow 1$ and  $\beta\rightarrow 1$, thus $A\rightarrow 1$ and
\begin{equation}
 F_x^{u-MGGA}\rightarrow F_x^{1} \rightarrow 1+\mu s^2 \; . 
\end{equation}
Note that in the limit of large atoms $\mu\rightarrow \mu^{MGE2}$, 
such that the semiclassical atom theory \cite{elliottPRL08,constantinPRL11} is 
correctly recovered.
\item[--] In the density tail asymptotic region  with
valence orbitals having a non-zero angular momentum quantum number
($s\rightarrow\infty, z< 1)$  so that 
$\beta \rightarrow 0$. We have also that $\redu\rightarrow \infty$
and $A\rightarrow 1$ (as $\beta^{(1/\redu)}\rightarrow 1$), thus 

\begin{equation}
F_x^{u-MGGA} \rightarrow  F_x^{1} \rightarrow \frac{2\pi\sqrt{\alpha}}{3\sqrt{5}\sqrt{\ln(1+\alpha)}}\ .
\label{ee8}
\end{equation}
Equation (\ref{ee8}) is an exact meta-GGA constraint for metallic surfaces \cite{constantin2016semilocal},
making asymtotically exact both the exchange energy per particle
and the potential, while for finite systems we found (see Appendix \ref{appb}) that the 
exchange energy per particle decays as $\epsilon_x\rightarrow -C/r^{3/2}$, and the exchange potential decays as
$v_x\rightarrow -C/(2 r^{3/2})$, with $C$ being a constant dependent on the angular momentum quantum number of the 
outer shell,  if $\alpha\rightarrow \infty$ \cite{dellasalaPRB15}.
Thus, for finite systems Eq. (\ref{ee8}) is not an exact constraint 
\cite{dellasalaPRB15,computation4020019}.
Nevertheless, this behavior is definitely more
realistic than the usual exponential decay behavior of
most semilocal functionals. 
In any case, we underline that  the Hartree potential is not used to describe the asymptotic region, 
in contrast to functionals based on the Fermi-Amaldi potential: instead, the meta-GGA expression in Eq. (\ref{ee8}) is used.
Moreover, in this work we are only considering non self-consistent results, which are thus quite unaffected by the 
choice of the functional in the asymptotic region. 

\end{itemize}
\end{itemize}


We note that the u-meta-GGA enhancement factor diverges for
$\redu\rightarrow+\infty$ and/or $s\rightarrow+\infty$ (i.e. in the tail of the density).
Therefore, unlike other functionals, it
does not respect the local form of the 
Lieb-Oxford bound \cite{liebIJQC81,constantinPRB15,feinblum2014communication}.
We note that this feature is anyway not an exact constraint and
it is indeed also strongly violated
by the conventional exact exchange energy density \cite{vilhenaPRA12}.
%
However, we recall that the global Lieb-Oxford bound, which is the true exact condition, 
is not tight, being usually fulfilled for all known physical systems, by most of the functionals \cite{odashimaJCP07}.  
As shown in the next section, the u-meta-GGA functional is accurate for atoms and molecules.
Thus, it implicitly 
satisfies the Lieb-Oxford bound for these systems.

\subsection{Parametrization of the functional}
\label{sec:param}
\begin{figure}
\includegraphics[width=\columnwidth]{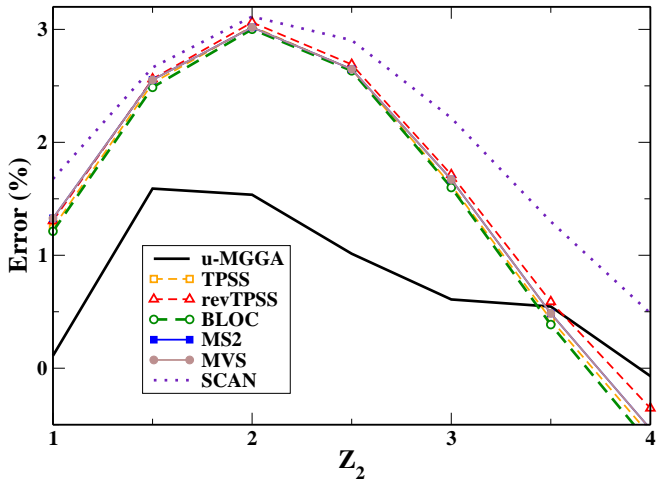}
\caption{Percent error
$(E_x^{exact}-E_x^{approx})/E_x^{exact}\times 100$ versus $Z_2$ for the model systems
described in Eq. (\ref{ebi2}).
}
\label{f3bb}
\end{figure}
Because the u-meta-GGA is exact for any one- and two-electron systems, 
we require it to be as accurate as possible also
for four-electron systems. To this purpose, 
we consider the four-electron hydrogenic-orbital model ($1s^22s^2$), with the 
following one-electron wavefunctions ($\psi_{nlm}$ with $n$, $l$, and $m$ being the principal, 
the angular, and the azimuthal 
quantum numbers respectively)
\begin{eqnarray}
& \psi_{100}(r)=\sqrt{\frac{1}{\pi}}Z_1^{3/2}e^{-Z_1r},\nonumber\\
& \psi_{200}(r)=\frac{1}{8}\sqrt{\frac{2}{\pi}}Z_2^{3/2}e^{-Z_2r/2}(2-Z_2r),
\label{ebi2}
\end{eqnarray}
with $Z_1$ and $Z_2$ being the nuclear charges seen by the $1s$ and $2s$ electrons, 
respectively. Note that for the real beryllium atom, $Z_1\approx 4$, and $Z_2\approx 2$. 
This model system is analytical and simple, and can 
cover important physics by varying $Z_1$ and $Z_2$. 
In Appendix \ref{appa} we show in detail the case $Z_1=Z_2=Z$. We 
also recall that the hydrogenic orbitals are important model systems in DFT, 
having been used to find various exact conditions 
\cite{heilmann1995electron,dellasalaPRB15,computation4020019} and 
to explain density behaviors 
\cite{heilmann1995electron,sunJCP12}.

%
%

The parameters $a_1$, $a_2$ and $a_3$ have been fitted by
fixing $Z_1=4$ (as in the beryllium case) and varying $Z_2$ between 1 and 4. 
In Fig. \ref{f3bb} we show the resulting percent error (i.e. 
$100\times(E_x^{exact}-E_x^{approx})/E_x^{exact}$) as a function of $Z_2$ and
we compare the u-meta-GGA results with those of other popular functionals. 
All the considered meta-GGA exchange functionals 
(TPSS \cite{taoPRL03}, revTPSS \cite{perdewPRL09}, BLOC \cite{constantinJCTC13}, 
MGGA\_MS2 \cite{sunPRL13,sunJCP13}, MVS \cite{sunPNAS15}, and SCAN \cite{sunPRL15}) 
perform similarly, while u-meta-GGA improves 
considerably, showing errors below 1.5 \%. 
%
%

\begin{figure}
\includegraphics[width=\columnwidth]{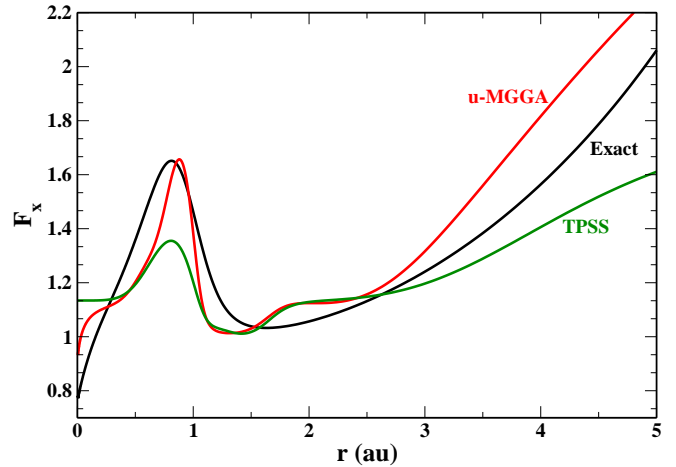}
\caption{Exchange enhancement factor $F_x$ versus the radial distance $r$, for Be atom.
}
\label{f3}
\end{figure}

%
In Fig. \ref{f3}, we report the u-meta-GGA exchange enhancement factor for the
Be atom, comparing it to the exact one (obtained as the ratio of the
conventional exact exchange and the LDA exchange energy densities) and
the popular TPSS meta-GGA.
This is a difficult and important example for the u-meta-GGA, because in the 
atomic core the density varies rapidly, the $1s$ and $2s$ orbitals overlap strongly, showing a significant amount of 
non-locality. Thus, $s$ 
and $\alpha$ are 
large ($s\approx 2$ at $r=0.8$, and $\alpha\approx 7$ and $r=1$), while $z$ is 
relatively small ($z\approx 0.34$ at $r=1$). See also Fig. \ref{fig1}.
$F_x^{u-MGGA}$ is smooth and more realistic 
than the TPSS one, at every point in space. Remarkably, the u-meta-GGA can also describe
well the atomic core.
Using the PBE \cite{perdewPRL96} orbitals and densities, the total exchange energies for Be atom are:
$E_x^{exact}=-2.659$ Ha, $E_x^{TPSS}=-2.673$ Ha, and $E_x^{u-MGGA}=-2.655$ Ha. 
%

\section{Computational Details}
All calculations for spherical systems (atoms, ions, and jellium clusters) have been performed with the 
numerical Engel code \cite{engel1993accurate,engel2003orbital}, using PBE orbitals and densities.

All calculations for molecules  have
been performed with the TURBOMOLE program package 
\cite{turbomole,furche2014turbomole} using PBE 
\cite{perdewPRL96} orbitals and densities and a def2-TZVPP basis set 
\cite{weigend2003gaussian,weigend2005balanced}. 
Similar results (not reported) have been found using LDA and 
Hartree-Fock orbitals and densities.

{Following a common procedure in DFT calculations, we have set a minimum
threshold ($10^{-20}$) for the electron density in order
to avoid divide-by-zero overflow errors in tail regions and one-electron
systems. All results are completely insensible to the value of the threshold.}

\section{Results}

\subsection{One- and two-electron systems}
\label{onedens_sec}
For one- and two-electron systems, the u-meta-GGA functional satisfies the exact condition in Eq. (\ref{exact}).
This is a very powerful exact constraint, that 
cannot be achived at the GGA and meta-GGA levels
of theory. In fact, even if some meta-GGAs have been fitted to the exchange
energies of the hydrogen atom (e.g. TPSS \cite{taoPRL03}, 
revTPSS \cite{perdewPRL09}, BLOC \cite{constantinJCTC13,constantinPRB13}, 
and Meta-VT\{8,4\} \cite{delCPL12}),
they are not exact for many other interesting one- and two-electron densities.
On the contrary, the u-meta-GGA functional is exact, not only for total exchange energies, but also for
exchange energy densities and potentials,
by construction, in all cases.

To make this point more clear, we consider briefly some relevant examples
of one- and two-electron densities. The first case concerns the
hydrogen (H), Gaussian (G), and cuspless hydrogen (C) one-electron
densities, that are defined as
\begin{equation}
n_H(r) = \frac{e^{-2r}}{\pi}\;\; ,\;\; n_G(r)=\frac{e^{-r^2}}{\pi^{3/2}}\;\; , \;\; n_C(r)=\frac{(1+r)e^{-r}}{32\pi}\ .
\end{equation}
These densities are models for atomic, bonding, and solid-state systems
\cite{taoPRL03,constantinPRB11,constantinJCP12}. They have
analytical exchange energies $E_H=-5/16$, $E_G=-1/\sqrt{2\pi}$,
and $E_C=-63/512$. Thus, we have used them to test the performance of
several functionals (see Table \ref{tab_one}).
\begin{table}
\caption{\label{tab_one}Relative errors ($10^3\times(E_x^{approx}-E_x^{exact})/E_x^{exact}$) for the exchaneg energy of 
the H, G, and C one-electron densities.}
\begin{ruledtabular}
\begin{tabular}{lrrr}
Functional & H & G & C \\
\hline
u-meta-GGA & 0.0 & 0.0 & 0.0 \\
TPSS & 0.0 & 0.3 & -3.6 \\
revTPSS & 0.0 & 1.5 & -3.3 \\
MS2 & 0.0 & -9.4 & -7.0 \\
MVS & 0.0 & -5.9 & -5.5 \\
SCAN & 0.0 & -3.5 & -4.8 \\
\end{tabular}
\end{ruledtabular}
\end{table}
Inspection of the table immediately shows that only the
u-meta-GGA is exact in all cases, whereas the other functionals
can at most perform exactly in a single case, by virtue of a
targeted parametrization. Note that any meta-GGA can not give the 
exact exchange potential of any one- or two- electron densities.

\begin{figure}
\includegraphics[width=0.95\columnwidth]{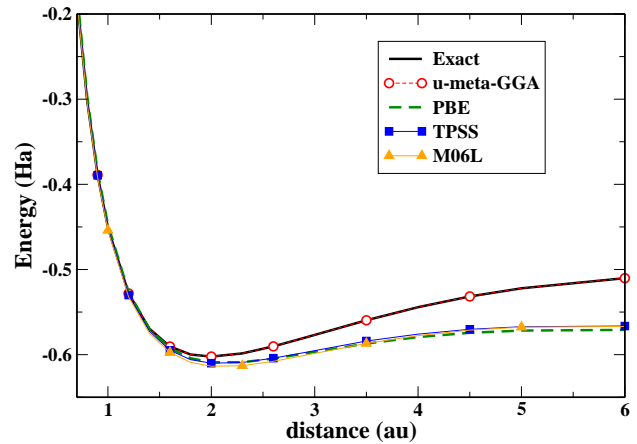}
\caption{\label{fig_h2}Dissociation curve of the H$_2^+$ molecule as computed with different functionals.}
\end{figure}

Another example is shown in Fig. \ref{fig_h2}, where we
plot the dissociation curve of the H$_2^+$ molecule, which is the simplest possible molecule.
This is a notoriously difficult problem for semilocal functionals \cite{cohen2008insights}, being related to the 
delocalization error.
Nevertheless, because the u-meta-GGA is exact for any one-electron density,
it yields the exact description for this difficult case.

\begin{figure}
\includegraphics[width=\columnwidth]{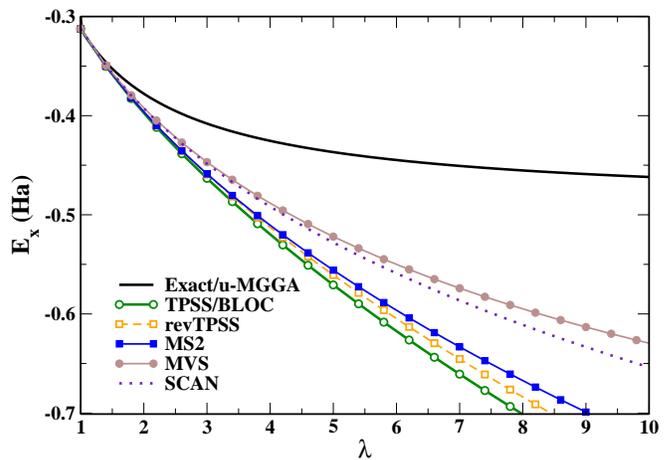}
\caption{Exchange energy (Ha) versus the scaling parameter $\lambda$, for
the non-uniformly scaled hydrogen atom in one direction \cite{kurth2000exchange}.}
\label{f4bb}
\end{figure}

Finally, we report in Fig. \ref{f4bb} the exchange energy 
computed for the non-uniformly scaled hydrogen atom versus
the scaling parameter $\lambda$ \cite{kurth2000exchange}. 
This is a model for quasi-two-dimensional systems and to study the
three-dimensional to two-dimensional crossover \cite{chiodoPRL12}.
All functionals, including meta-GGAs, are very accurate at $\lambda=1$ (i.e.
the conventional three-dimensional hydrogen atom). However, for larger
values of the confining parameter only u-meta-GGA is exact (by construction).
The meta-GGA functionals instead fail badly even for mild and moderately 
large values of $\lambda$.

Other examples of two-electron densities of interest in DFT are the
Hooke's atom \cite{taut94,constantinJCTC13,sun16}, the Loos-Gill model 
\cite{sun16,loosgill}, and the strictly-correlated two-electrons model 
\cite{seidl16,buttazzo12}. In all these cases, the
u-meta-GGA functional yields, by construction, an exact description of exchange.

\subsection{Atoms}
Computing the absolute energies of atoms can be expected to be
quite a hard task for the u-meta-GGA functional.
In fact, the functional is exact for one- and two-electron system
(i.e. H and He atoms) but
for increasingly large atoms the Hartree reduced parameter 
becomes soon very large (see Fig. \ref{fig1}).
Therefore, a  particular care is required to balance the 
contribution of this ingredient in different cases.

\begin{figure}
\includegraphics[width=\columnwidth]{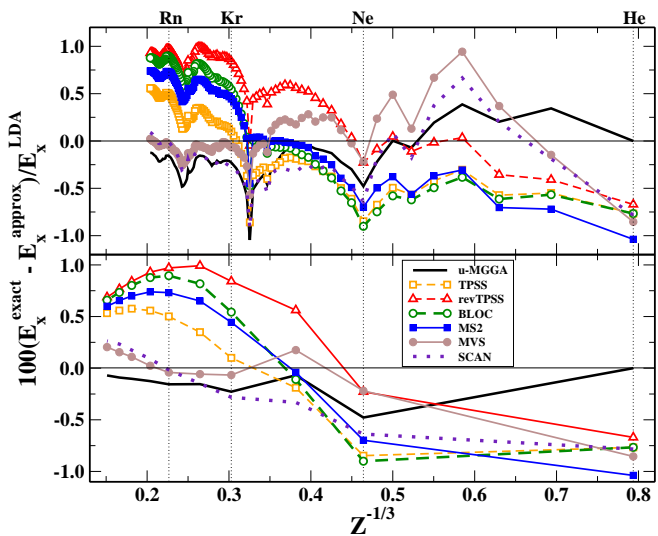}
\caption{ Upper panel: Percent exchange energy error ($100(E_x^{exact}-E_x^{approx})/E_x^{LDA}$) versus $Z^{-1/3}$ for all periodic table atoms ($2\leq Z\leq 118$). Lower panel: Percent exchange energy error ($100(E_x^{exact}-E_x^{approx})/E_x^{LDA}$) versus $Z^{-1/3}$, for noble atoms ($2\leq Z\leq 290$).}
\label{f5bat}
\end{figure}

To check this issue, we have calculated the
exchange energy of all periodic table atoms ($2\leq Z\leq 118$) and we have compared
the u-meta-GGA results to those of some meta-GGA functionals.
The results are reported in the upper panel of Fig. \ref{f5bat} and in
Table \ref{tab1}.
\begin{table*}[bt]
\caption{\label{tab1}
Mean absolute errors (mHa) for various systems and properties.
}
\begin{ruledtabular}
\begin{tabular}{llrrrrrrr}
System & Property & TPSS & revTPSS & BLOC & MS2 & MVS & SCAN & u-MGGA  \\
\hline
Atoms ($2\leq Z\leq 118$) & $E_x/Z$ & 12.4 & 27.3 & 23.4 & 18.9 & \bf{2.8} & 4.8 & 7.6 \\
Noble atoms ($2\leq Z \leq 290$) & $E_x/Z$ & 21.5 & 34.2 & 31.6 & 27.2 & 5.5 & 8.8 & \bf{5.1} \\
4$e^-$-ions ($4\leq Z\leq 20$) & $E_x$  & 50.5 & 29.8 & 56.8 & 51.1 & 65.1 & 37.6 & \bf{2.5} \\
7$e^-$-ions ($7\leq Z\leq 23$) & $E_x$  & 23.3 & 64.9 & 25.0 & 13.7 & 68.8 & \bf{4.3} & 8.6 \\
10$e^-$-ions ($10\leq Z\leq 26$) & $E_x$  & 30.6 & 156.0 & 35.4 & 39.7 & 59.3 & 76.9 & \bf{12.2} \\
29$e^-$-ions ($29\leq Z\leq 45$) & $E_x$  & 203.8 & 754.0 & 483.5 & 448.2 & \bf{63.4} & 225.8 & 191.1 \\
Jellium clsusters $r_s=4$ ($2\leq Z\leq 92$) & $E_x/Z$ & 1.1 & 1.3 & 1.1 & \bf{0.9} & 2.0 & 1.3 & 1.0 \\
jellium clsusters $r_s=1$ ($2\leq Z\leq 92$) & $E_x/Z$ & 2.7 & 4.7 & 3.4 & 3.8 & 2.7 & \bf{1.7} & 2.1 \\
\end{tabular}
\end{ruledtabular}
\end{table*}
%

%
%
%

The u-meta-GGA performs remarkably well for all the periodic table atoms, being one of the most accurate 
functionals, with a mean absolute error (MAE) of 7.6 mHa/electron, slightly worse than 
MVS and SCAN meta-GGAs (with MAE=2.8 mHa/electron and MAE=4.8 mHa/electron, respectively).
Moreover, in Fig. \ref{f5bat}, we show the exchange energy error 
($100(E_x^{exact}-E_x^{approx})/E_x^{LDA}$) for
noble atoms with $2\leq Z\leq 290$. This plot shows that, 
in case of large atoms ($118\leq Z\leq 290$), 
the u-meta-GGA becomes the most accurate functional, due to the semiclassical 
atom theory which it incorporates. 

\subsection{Isoelectronic series and Jellium clusters}
\begin{figure*}
\includegraphics[width=\textwidth]{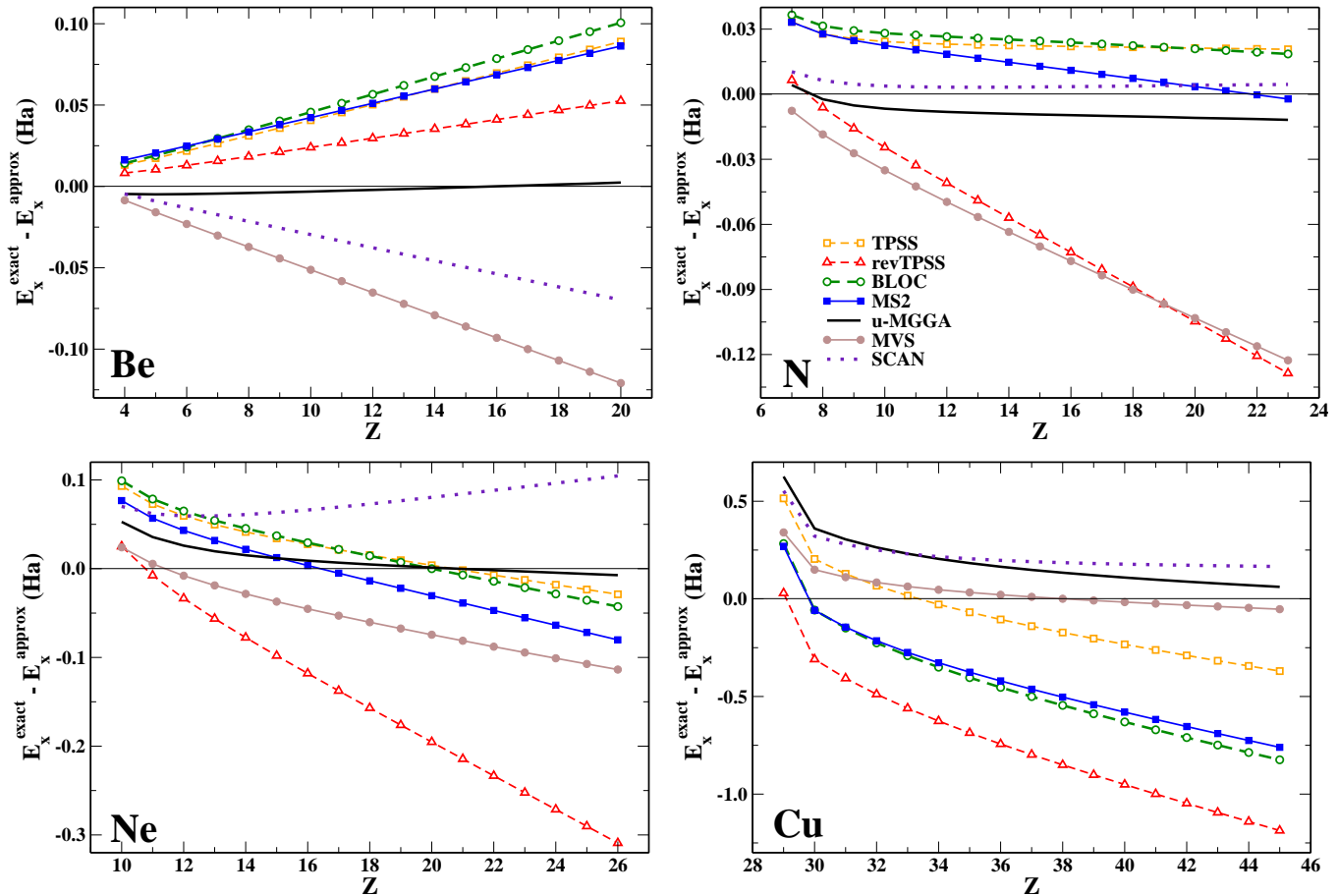}
\caption{Exchange errors ($E_x^{exact}-E_x^{approx}$) versus nuclear charge
  $Z$, for Beryllium (top-left panel), Nitrogen (top-right panel), Neon (bottom-left panel), and Copper (bottom-right panel)
isoelectronic series.
}
\label{fiso}
\end{figure*}

We consider the first 17 ions of the isoelectronic series of 
Beryllium ($4\leq Z \leq 20$), Nitrogen ($7\leq Z \leq 23$), 
Neon ($10\leq Z \leq 26$), and Copper ($29\leq Z\leq 45$). 
The results for all systems are reported in Fig. \ref{fiso} while 
the MAEs are shown in Table \ref{tab1}. 
The u-meta-GGA is very accurate in all cases, outperforming most of the other semilocal functionals for Be and Ne. 
For N (Cu) the best functional is SCAN (MVS) and u-meta-GGA is the second best one.
Note that the case of Cu is the most difficult one, because u-meta-GGA performs modestly for the Cu atom (see Fig. \ref{f5bat}). 
Nevertheless,  for increasing $Z$ values it soon becomes very accurate.

%
We also tested the u-meta-GGA for magic jellium clusters with
2, 8, 18, 20, 34, 40, 58, and 92 electrons for
bulk parameters $r_s=1$ and $r_s=4$. The error statistics are reported in 
Table \ref{tab1}. In both cases u-meta-GGA is accurate, being in line with the 
best semilocal functionals.

\subsection{Molecules}
In Table \ref{tab2} we report the exchange atomization energies
of the systems constituting the AE6 test set \cite{lynch2003small},
as computed with several methods.
\begin{table*}[bt]
\begin{center}
\caption{\label{tab2}
Errors (kcal/mol) and error statistics for the exchange atomization energies of the AE6 test. The best result of each line is highlighted in bold style.}
\begin{ruledtabular}
\begin{tabular}{lrrrrrrr}
 & TPSS & revTPSS & BLOC & MS2 & MVS & SCAN & u-MGGA  \\
\hline
CH$_4$ & -3.1 & \textbf{-1.9} & -2.8 & -3.0 & 3.8 & 5.6 & -15.6 \\
SiO & 31.8 & 30.5 & 27.8 & 26.7 & 38.4 & 34.3 & \textbf{12.9} \\
S$_2$ & 17.4 & 18.2 & 15.1 & 14.4 & 25.1 & 14.3 & \textbf{3.6} \\
C$_3$H$_4$ & 20.2  & 15.2 & 16.0 & 25.4 & 42.1 & 45.5 & -\textbf{1.3} \\
C$_2$H$_2$O$_2$ & 60.2 & 56.9 & 51.9 & 66.0 & 78.2 & 83.6 & \textbf{16.3} \\ 
C$_4$H$_8$ & \textbf{-2.8} & -8.5 & -11.2 & 21.3 & 34.2 & 48.4 & -47.6 \\ 
\hline
MAE & 22.6 & 21.9 & 20.8 & 26.1 & 37.0 & 38.6 & \textbf{16.2} \\   
MARE & 13.8 & 13.6 & 12.1 & 12.8 & 19.4 & 16.0 & \textbf{5.7} \\
\end{tabular}
\end{ruledtabular}
\end{center}
\end{table*}
One can see that the u-meta-GGA functional performs
quite well in this case, being often superior
to meta-GGA functionals and yielding overall
the best MAE.
This result shows that the u-meta-GGA functional
provides a well balanced description of atoms and molecules, at the exchange level.
We note that this success goes beyond the exactness of this functional
for one- and two-electron systems, since in the present case
this feature concerns only the computation of the H atom energy,
which is exact also for all the other tested meta-GGAs.

As additional test, we consider in Table \ref{tab3}
the exchange-only barrier heights and reaction energies 
of the systems defining the K9 test set \cite{lynch2003robust}.
This is a harder test than the previous one, since transition-state
structures display rather distorted geometries and are therefore 
characterized by a different density regime than ordinary molecules.
\begin{table*}[tb]
\begin{center}
\caption{ \label{tab3}
Errors (kcal/mol) and error statistics of several exchange functionals for the K9 representative test. The best result 
of each line is highlighted in bold style.}
\begin{ruledtabular}
 \begin{tabular}{lrrrrrrr}
 System & TPSS & revTPSS & BLOC & MS2 & MVS & SCAN & u-MGGA \\
 \hline
 \multicolumn{8}{c}{Forward barriers}\\
 OH+CH$_4\rightarrow$CH$_3$+H$_2$O & -12.5  & -11.6  & -10.5  & -10.2 & -11.1 & -12.1 & \textbf{-1.0} \\
  H+OH$\rightarrow$O+H$_2$ & -2.7   & -8.1  & -6.8   & \textbf{-2.2} & -5.5 & -3.0 & -8.4  \\
 H+H$_2$S$\rightarrow$H$_2$+HS & \textbf{-1.0}   & -3.6   & -4.1   & -3.4  & -4.5 & -5.0 & -3.1 \\
MAE & 8.0 & 7.8 & 7.1 & 5.3 & 7.0 & 6.7 & \bf{4.2} \\
 & & & & & & & \\
 \multicolumn{8}{c}{Backward barriers}\\
OH+CH$_4\leftarrow$CH$_3$+H$_2$O & -7.6   & -8.1 & -6.8 & -10.3 & 2.1 & -5.5 & \textbf{3.6} \\
 H+OH$\leftarrow$O+H$_2$ & -10.4  & \textbf{-9.9}    &  -12.2   & -15.7 & -13.6 & -16.6 & -16.9 \\
 H+H$_2$S$\leftarrow$H$_2$+HS & -2.2   & \textbf{-0.9}  & -1.6  & -8.2 & -2.9 & -7.0 & -3.7 \\
MAE & 7.3 & \textbf{6.3} & 6.9 & 8.9 & 8.7 & 9.7 & 8.1 \\
 & & & & & & & \\
 \multicolumn{8}{c}{Reaction energies}\\ 
 $\Delta$(OH+CH$_4$-CH$_3$+H$_2$O) & -4.9  & -3.5  & -3.7  &  \textbf{0.1} & -13.2 & -6.5 & -4.6 \\
 $\Delta$(H+OH-O+H$_2$) & 7.7  & \textbf{1.8}  & 5.3  & 11.4 & 10.2 & 13.6 & 8.5 \\
 $\Delta$(H+H$_2$S-H$_2$+HS) & -2.0  & -2.7  & -2.5  & -0.5 & 3.7 & 2.1 & \textbf{0.6} \\
MAE & 3.9 & \bf{2.7} & 3.8 & 4.0 & 9.0 & 7.4 & 4.6 \\
 & & & & & & & \\
 \multicolumn{8}{c}{Overall statistics}\\
MAE & 6.4 & \bf{5.6} & 6.0 & 6.1 & 8.3 & 7.9 & \textbf{5.6} \\
 \end{tabular}
\end{ruledtabular}
 \end{center}
 \end{table*}
Inspection of the table shows  that the errors on reaction energies
display a similar trend as for the atomization energies, even though the
differences between the functionals are smaller because of the 
smaller magnitude of the computed energies.
Instead, for barrier heights no clear trend can be extracted.
Nevertheless, the u-meta-GGA functional shows a reasonable performance
being similar to meta-GGAs. 
This finding supports the robustness of the construction
presented in Section \ref{umgga_sec}.

\section{Compatibility of the u-meta-GGA with semilocal correlation functionals}
In this section we investigate the possibility to combine the
u-meta-GGA exchange with an existing semilocal correlation functional.
Thus, we consider the performance of  different combinations
of the u-meta-GGA exchange with an existing semilocal correlation functional,
for the description of molecular properties, namely the
AE6 \cite{lynch2003small,Haunschild2012} and K9 \cite{lynch2003robust,Haunschild2012} 
test sets.
In more detail, we consider the following correlation functionals: 
PBE \cite{perdewPRL96}, PBEloc \cite{constantinPRB12}, GAPloc \cite{fabianoJCTC14gap}, TCA \cite{tognettiCPL08}, vPBE \cite{sunJCP13} (the semilocal correlation of the MGGA-MS functional), PBEsol \cite{perdewPRL08}, LYP \cite{leePRB88} [GGA functionals], 
TPSS \cite{taoPRL03}, revTPSS \cite{perdewPRL09,perdewPRL11}, 
BLOC \cite{constantinJCTC13}, JS \cite{constantinPRB11js} 
[meta-GGA functionals].  

\begin{figure}
\includegraphics[width=\columnwidth]{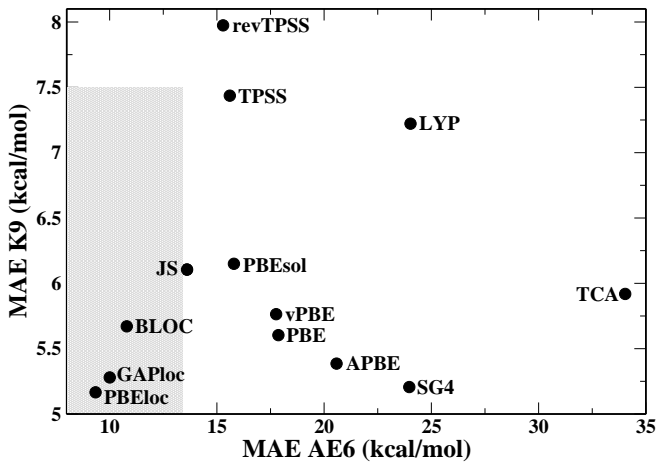}
\caption{Mean absolute error (MAE) on the AE6 test versus MAE on the K9 test
  for the combination of the u-meta-GGA exchange with different semilocal
  correlation functionals. The grey-shaded area highlights the combinations
  that perform better than the PBE XC functional.}
\label{f7}
\end{figure}

In Fig. \ref{f7} we report the MAE on the AE6 test versus the
MAE for the K9 test as obtained by the different functionals.
%
%
The best performance is found for 
PBEloc, GAPloc, and BLOC. These are indeed the
only correlation functionals that allow to achieve for both tests
results that are better than the simple PBE XC ones
(13.4 kcal/mol for AE6 and 7.5 kcal/mol for K9), which
we have used here as a reference.
This result indicates that
a more localized correlation energy density may favor
the compatibility with the u-meta-GGA in finite systems.
This conclusion can be traced back to the fact that
the localization constraint in the PBEloc, GAPloc, and BLOC
correlation functionals has been introduced to 
enhance the compatibility of the semilocal correlation with
exact exchange \cite{constantinPRB12,fabianoJCTC14gap}, thus it
also improves the compatibility with the u-meta-GGA exchange which
is rather close to the exact one.

Nevertheless, we find that none of the semilocal correlation functionals
can yield highly accurate results, when used with the u-meta-GGA exchange.
{This is not much surprising since the usual error cancellation
that occurs at the semilocal level between exchange and correlation
contributions cannot work properly in this case because the u-meta-GGA
functional is exact for one- and two-electron systems.}
This suggests 
the need for the construction of a proper
u-meta-GGA correlation functional being able to include the non-local
effects on equal footing with the exchange part. Such a development 
is anyway not trivial, since it requires the 
development of a highly accurate correlation functional for 
two-electron systems, {including also static correlation effects, that
are (correctly) not accounted for by the u-meta-GGA exchange (in contrast to
simple semilocal exchange functionals).
Such functionals are usually developed at the hyper-GGA level of theory
\cite{perdewPRA08,haunschildJCP12,beckeJCP07,becke05}
and they include exact exchange as a basic input ingredient. However,
the use of exact exchange as an ingredient would make the u-meta-GGA
construction of the exchange term meaningless}. 
{A possible strategy to solve this dilemma can be to consider a smooth interpolation
of a hyper-GGA expression for one- and two-electron cases (where $z=1$
and the exact exchange is given by the Hartree potential) with a more traditional
semilocal correlation expression for many-electron cases.}
{Anyway, this very challenging task}
will be
the subject of other work.


\section{Conclusions}
The success of semilocal DFT is mainly based on the correctness of the semiclassical physics that 
it incorporates (e.g. gradient expansions derived from small perturbations of the uniform electron gas), and on 
the satisfaction of several formal exact properties (e.g. density scaling relations).
However, it also relays on a heavy error cancellation 
between the exchange and correlation parts. Thus, semilocal DFT can often achieve good accuracy for large systems, 
where the semiclassical physics is relevant, but not for small systems, that are usually 
treated with hybrid functionals. 

Using the reduced Hartree parameter $\redu(\R)$ [Eq. (\ref{e6})] as a new
ingredient in the construction of DFT functionals, can guarantee the exactness
of the exchange functional for any one- and two-electron systems. This is an
important exact condition, also related to the homogeneous density scaling
\cite{scaling,borgooJCP12,borgooJCTC13}, the delocalization and many-electron
self-interaction errors \cite{cohen2008insights}, and it can boost the
accuracy of the functional. 

Hence, we have constructed a prototype u-meta-GGA exchange functional, showing that it is 
possible and useful the use of the reduced Hartree parameter $\redu(\R)$. 
Note that even if $\redu(\R)$ is non-local,
we have shown that it is compatible with the semilocal quantities. The u-meta-GGA has been
tested for a broad range of finite systems (e.g. atoms, ions, jellium spheres,
and molecules)
being better than, or comparable with, the popular meta-GGA exchange functionals.

Nevertheless, we have showed that $\redu(\R)$ is a
size-extensive quantity, increasing with the number of electrons. This fact
represents a real challenge for functional development, 
{limiting the applicability of the present formalism to periodic
  (infinite) systems}. This limitation can be removed
only by a large screening. Such a screening is given, in the present work, [
Eqs. (\ref{e7})-(\ref{e9})] by the function $\beta(s,z)$. An alternative way will be
the use of the screened reduced Hartree potential $x^u(\R)$
\cite{computation4020019}, defined by 
\begin{equation}
\sredu(\R)=\frac{1}{3(3n(\R)/\pi)^{1/3}}\int d\R' \frac{n(\R')}{|\R-\R'|}e^{-a
\alpha(\R')^b k_F(\R')^\beta|\R-\R'|^{\beta}},
\label{eq18}
\end{equation}
where $a$, $b$, and $\beta$ are other positive constants. Note that $\sredu(\R)=\redu(\R)$ 
for any one- and two-electron systems, 
and $\sredu(\R)$ is realistic at the nuclear region \cite{computation4020019}. 
However, such an approach, which is theoretically more powerful, 
is significantly more complex. In addition, it is also
computationally more expensive since the bare Hartree potential 
is computed at every step of the Kohn-Sham 
self-consistent method, and thus its use does not affect the speed of the calculation,
whereas the screened Hartree potential should be calculated separately for
the only purpose of constructing the functional.    

We also note that the bounded ingredient $\redub$ of Eq. (\ref{ebi}), can by itself be of 
interest for the development of exchange-correlation and even kinetic functionals,
since it is a powerfull atomic indicator. 
In this sense, a further investigation of 
this issue may be worth. Construction of the exchange enhancement factors 
of the form $F_x(s,\redub)$ should be much simpler, 
because $\redub$ is bounded, and should reveal the importance of 
the non-locality contained in this ingredient. 

In any case, the  u-meta-GGA exchange functional defined in Eqs. (\ref{e7})-(\ref{eqlast}) is just a first attempt,
and other simpler and/or better functional forms could possibly be developed.
Thus, the class of u-meta-GGA functionals may represent a new semi-rung on the Jacob's ladder: it is  above the third one 
as it includes the Hartree potential to describe exactly the exchange for any one- and two-electron systems, but with a 
computational cost lower than functionals dependent on exact exchange.
In this work, all calculations are non-self consistent. In a future work we will consider the functional derivative of
the u-meta-GGA functionals.
%

\textbf{Acknowledgments.} We thank TURBOMOLE GmbH for the TURBOMOLE program package.

\appendix

\section{Size consistency}
\label{appsize}
%
%
Because the Hartree reduced parameter $\redu$ is a size extensive quantity,
it is important to prove that the u-meta-GGA functional
is properly size consistent.
%
That is, given two systems, 
$A$ and $B$, separate by an infinite distance and whose densities are
not overlapping, we have 
\begin{equation}\label{e10}
E_x^{u-MGGA}[A+B]=E_x^{u-MGGA}[A]+E_x^{u-MGGA}[B]\ ,
\end{equation}
where $E_x^{u-MGGA}=\int n\epsilon_x^{LDA}F_x^{u-MGGA}d\R$.
To show this, we can use the fact that the integrand is finite everywhere 
($n\epsilon_x^{LDA}$ decays exponentially, while in the evanescent density regions 
$F_x^{u-MGGA}$ behaves according to Eq. (\ref{ee8})), to write
\begin{eqnarray}\label{e11}
& E_x^{u-MGGA}[A+B]=\int_{\Omega_A}n\epsilon_x^{LDA}F_x^{u-MGGA}d\R + 
\nonumber \\
& \int_{\Omega_B}n\epsilon_x^{LDA}F_x^{u-MGGA}d\R\ ,
\end{eqnarray}
where $\Omega_A$ and $\Omega_B$ are the space domains where 
$n_A$ and $n_B$, respectively, are not zero.
Then, considering any $\R\in\Omega_A$ (analogous
considerations hold for $\Omega_B$), we have
\begin{eqnarray}
\label{e12}
\redu_{A+B}(\R) & = & \frac{\int\frac{n_A(\R')+n_B(\R')}{|\R-\R'|}d\R'}{3(3/\pi)^{1/3}(n_A(\R)+n_B(\R))^{1/3}} \\
\nonumber
& = & \frac{\int_{\Omega_A}\frac{n_A(\R')}{|\R-\R'|}d\R'+\int_{\Omega_B}\frac{n_B(\R')}
{|\R-\R'|}d\R'}{3(3/\pi)^{1/3}(n_A(\R)+n_B(\R))^{1/3}}\ .
\end{eqnarray}
Now, because $\R\in\Omega_A$, we have that $n_B(\R)=0$; moreover,
because the two systems lay at infinite distance from each other, 
$|\R-\R'|=\infty$ for any $\R'\in\Omega_B$.
Hence,
\begin{equation}\label{e13}
\redu_{A+B}(\R) = \frac{\int_{\Omega_A}\frac{n_A(\R')}{|\R-\R'|}d\R'+}{3(3/\pi)^{1/3}(n_A(\R))^{1/3}} 
= \redu_A(\R)\ .
\end{equation}
In the same way, for $\R\in\Omega_B$ we have $\redu_{A+B}(\R) = \redu_B(\R)$.
At this point, since all the other input quantities are semilocal,
Eq. (\ref{e11}) immediately yields Eq. (\ref{e10}).

\section{Asymptotic behavior}
\label{appb}
In case of spherical systems in a central potential (e.g. atoms, jellium spheres), the following equation holds
\cite{dellasalaPRB15}
\begin{equation}
\tau-\tau^W=\frac{l(l+1)}{2}\frac{n}{r^2},
\label{eA1}
\end{equation}
in the asymptotic region. 
Here $l$ is the angular momentum quantum number of the outer shell, and the density decays
exponentially  $n\sim e^{-b r}$, when the radial distance is large ($r\rightarrow\infty$). Here $b=2\sqrt{-2\mu}$, with
$\mu$ being the ionization potential. Then,
for any $l\ne 0$, $\alpha$ diverges as
\begin{equation}
\alpha=\frac{l(l+1)}{2C_sn^{2/3}}\frac{1}{r^2},
\label{eA2}
\end{equation}
where $C_s=\frac{3}{10}(3\pi^2)^{2/3}$. Considering the enhancement factor of Eq. (\ref{ee8}),
i.e. $F_x^{MGGA}(\alpha)$, the exchange energy per particle
\begin{equation}
\epsilon_x=-C_x n^{1/3}F_x^{MGGA}(\alpha) \;\;\; \mathrm{with}\;\;\;  C_x=\frac{3}{4}(\frac{3}{\pi})^{1/3},
\label{eA3}
\end{equation}
decays as
\begin{equation}
\epsilon_x \rightarrow
-\frac{\sqrt{2}}{4}\frac{\sqrt{l(l+1)}}{\sqrt{b}}\frac{1}{r^{3/2}}+\mathcal{O}(\frac{1}{r^{5/2}}).
\label{eA4}
\end{equation}
Concerning the exchange potential, we consider the generalized Kohn-Sham framework
to write \cite{arbuznikov2002validation,dellasalaPRB15}
\begin{eqnarray}
\displaystyle v_x \phi_i &=&
%
\frac{\partial (n \epsilon_x)}{\partial n}\phi_i
 -\nabla\cdot  \left [ \frac{\partial (n \epsilon_x)}{\partial \nabla n} \phi_i
+\frac{1}{2} \frac{\partial (n \epsilon_x)}{\partial \tau} \nabla\phi_i
\right ] \nonumber  \\
&+&  \left(
 \frac{\partial (n \epsilon_x)}{\partial  \nabla n }
\right)\cdot\nabla\phi_i  \, .
\label{eA5}
\end{eqnarray}
Using the following equations
\begin{eqnarray}
&& \displaystyle
\frac{\partial (n\epsilon_x)}{\partial n}=-C_x\frac{4}{3}n^{1/3}F_x^{MGGA}(\alpha)-C_x n^{4/3}\frac{d
F_x^{MGGA}(\alpha)}{d\alpha}\frac{\partial\alpha}{\partial n},\nonumber\\
&& \displaystyle \frac{\partial (n\epsilon_x)}{\partial \nabla n}=-C_x n^{4/3}\frac{d F_x^{MGGA}(\alpha)}{d\alpha}
\frac{\partial\alpha}{\partial \nabla n},\nonumber\\
&& \displaystyle \frac{\partial (n\epsilon_x)}{\partial \tau}=-C_x n^{4/3}\frac{d F_x^{MGGA}(\alpha)}{d\alpha}
\frac{\partial\alpha}{\partial\tau},\nonumber\\
&& \displaystyle \frac{\partial\alpha}{\partial n}=\frac{|\nabla n|^2}{8 n^2
\tau^{unif}}-\frac{5}{3}\frac{\alpha}{n},\nonumber\\
&& \displaystyle \frac{\partial\alpha}{\partial \nabla n}=-\frac{\nabla n}{4 n\tau^{unif}},\nonumber\\
&& \displaystyle \frac{\partial\alpha}{\partial\tau}=\frac{1}{\tau^{unif}},\nonumber\\
&& \displaystyle \tau^{unif}=C_s n^{5/3},
\label{eA6}
\end{eqnarray}
we obtain after some simple algebra
\begin{eqnarray}
&& \displaystyle
v_x \phi=-C_x n^{1/3}(\frac{4}{3}F_x^{MGGA}-\frac{5}{3}\frac{d F_x^{MGGA}(\alpha)}{d\alpha}\alpha)\phi- \nonumber\\
&& \displaystyle
C_x n^{4/3}\frac{d F_x^{MGGA}(\alpha)}{d\alpha}\frac{\nabla n}{4 n\tau^{unif}}\cdot (\frac{\nabla
n}{2n}\phi-\nabla\phi)-\nonumber\\
&& \displaystyle
\nabla\cdot [C_x n^{4/3}\frac{d F_x^{MGGA}(\alpha)}{d\alpha}\frac{1}{2\tau^{unif}}(\frac{\nabla
n}{2n}\phi-\nabla\phi)],
\label{eA7}
\end{eqnarray}
where $\phi$ is the highest occupied orbital. Then, the asymptotic density is $n=f \phi^2$
(with $f$ being the occupation number) and
\begin{equation}
\frac{\nabla n}{2n}\phi-\nabla\phi=0.
\label{eA8}
\end{equation}
The final formula for the exchange potential is
\begin{equation}
v_x =-C_x n^{1/3}\left(\frac{4}{3}F_x^{MGGA}-\frac{5}{3}\frac{d F_x^{MGGA}(\alpha)}{d\alpha}\alpha\right),
\label{eA9}
\end{equation}
which is valid for any exchange enhancement factor that depends only on the $\alpha$ ingredient.
Then, the exchange potential of $F_x^{MGGA}$ defined in Eq. (\ref{ee8}) behaves at $r\rightarrow\infty$ as
\begin{equation}
v_x\rightarrow -\frac{\sqrt{2}}{8}\frac{\sqrt{l(l+1)}}{\sqrt{b}}\frac{1}{r^{3/2}}+\mathcal{O}(\frac{1}{r^{5/2}}).
\label{eA10}
\end{equation}

\section{Hydrogenic orbitals}
\label{appa}

The system of Eq. (\ref{ebi2}), with $Z_1=Z_2=Z$ has the following density
\begin{equation}
n(r)=2\,{\frac {{Z}^{3} \left( {{\rm e}^{-r\,Z}} \right) ^{2}}{\pi }}+
\,{\frac {{Z}^{3} \left( {{\rm e}^{-1/2\,r\,Z}} \right) ^{2} \left( 2-
r\,Z \right) ^{2}}{16\pi }},
\label{eeA1}
\end{equation}
kinetic energy density
\begin{equation}
\tau(r)={\frac {{Z}^{5} \left( {{\rm e}^{-r\,Z}} \right) ^{2}}{\pi }}+{\frac {
1}{128}}\,{\frac {{Z}^{5}{{\rm e}^{-r\,Z}} \left( -4+r\,Z \right) ^{2}
}{\pi }},
\label{eeA2}
\end{equation}
Hartree potential
\begin{eqnarray}
&& u(r)=\frac{1}{4r}(16-6ve^{-v}-8e^{-v}-2v^2e^{-v}-\nonumber\\
&& 8ve^{-2v}-8e^{-2v}-v^3e^{-v}),
\label{eeA3}
\end{eqnarray}
and exchange energy density
\begin{eqnarray}
&& e_x(r)=n(r)\epsilon_x(r)=-\frac{Z^3}{6912\pi r}(864 e^{-v}-864ve^{-v}+  \nonumber\\
&& 6048e^{-2v}+216v^2e^{-v}+216v^2e^{-2v}+1024v^2e^{-3v}- \nonumber\\
&& 768v^3e^{-3v}-6912e^{-4v}-6912ve^{-4v}+1024ve^{-3v}+ \nonumber\\
&& 216ve^{-2v}-27v^5e^{-2v}-54v^3e^{-2v}+54v^4e^{-2v}),
\label{eeA4}
\end{eqnarray}
where $v=Zr$. The Hartree, exact exchange and LDA exchange energies are 
\begin{eqnarray}
&& U=\frac{49565}{20736}Z=2.39029 Z,\nonumber\\
&& E_x=-\frac{305797}{373248}Z=-0.8192864 Z,\nonumber\\
&& E_x^{LDA}=-0.7183437428 Z.
\label{eeA5}
\end{eqnarray}
Note that all the exchange ingredients ($s$, $z$, $\alpha$, $\redu$) are only functions of $v=Zr$,
such that for any exchange enhancement factor $F_x(s,\alpha,z,\redu)$, the total exchange energy will be
$E_x=-constant\; Z$. 


\bibliography{umeta}
\end{document}